# The One Dimensional Hydrogen Atom Revisited

Guillermo Palma [*] and Ulrich Raff

*Department of Physics*
*9170019 Avenida Ecuador 3493*
*Casilla 307 Correo-2*
*University of Santiago de Chile*
*Santiago de Chile*

## Abstract

The one dimensional Schrödinger hydrogen atom is an interesting mathematical and physical problem to study bound states, eigenfunctions and quantum degeneracy issues. This 1D physical system gave rise to some intriguing controversy over more than four decades. Presently, still no definite consensus seems to have been reached. We reanalyzed this apparently controversial problem, approaching it from a Fourier transform representation method combined with some fundamental (basic) ideas found in self-adjoint extensions of symmetric operators. In disagreement with some previous claims, we found that the complete Balmer energy spectrum is obtained together with an odd parity set of eigenfunctions. Closed form solutions in both coordinate and momentum spaces were obtained. No twofold degeneracy was observed as predicted by the degeneracy theorem in one dimension, though it does not necessarily have to hold for potentials with singularities. No ground state with infinite energy exists since the corresponding eigenfunction does not satisfy the Schrödinger equation at the origin.

## 1. Introduction

One-dimensional quantum mechanical problems can be quite challenging and more complex from a topological point of view than their 3D counterpart due to the presence of singularities at the origin. This is well illustrated with the *one dimensional hydrogen atom* which has been solved and discussed using a large variety of approaches over more than four decades [1-17]. This problem has its own relevance since one can study bound states, degeneracy issues, singularity of the potential at the origin and the corresponding eigenfunctions. At a first glance, one could be under the impression that solving this problem, i.e. an electron moving in a one dimensional potential $-e/|x|$, is merely an exercise utilized to illustrate yet another example of introductory quantum mechanics. A closer look however, reveals that no final consensus has been reached so far which motivated our analysis. We found that our solution to this problem unambiguously allows clarification of all physical relevant claims and results obtained in the past. The controversies and confusions encountered in this particular problem first call for a careful look at previous works to hopefully shed some final light on the Schrödinger 1D hydrogen atom.

A standard treatment of this problem goes back as far as five decades [1] where only wave functions of odd parity were derived, i.e. those functions that change sign when reflecting $x \rightarrow -x$. No further in-depth analysis was provided. According to Loudon [2], these functions do not form a complete set since a priori no even function can be expanded in terms of them. In addition the wave function of the lowest bound state of one-.dimensional problems should be nodeless contradicting the original approach [1]. Loudon then re-examined the problem and came to the conclusion that the 1D hydrogen atom has a ground state with infinite binding energy while all the excited bound states have a twofold degeneracy having an even and odd wave function for each eigenvalue, apparently in contradiction with the theorem stipulating that 1D systems cannot have degenerate quantum states.

Loudon's results were partially corroborated by Andrews [3] who analyzed singular potentials in one dimension. On one hand Andrews concluded that Loudon's solutions are correct apart from the existence of a ground state [4]. On the other hand, Andrews argued that the singularity acts as an impenetrable barrier if the potential is not integrable up to the singularity and consequently finds no reason to match the solution on the right side of the singularity with the one on the left side. Andrews also rejects Haines and Roberts' results which consist of a physically incorrect continuum of negative energy eigenstates [5]. Even parity solutions, rejected by Haines and Roberts on the ground that they do not satisfy the Schrödinger equation at $x = 0$, were nevertheless accepted by Andrews arguing that since there is an impenetrable barrier at $x = 0$, adding a $\delta$ function to it, makes no difference.

A different approach was taken by Nieto [6] who solved the N-dimensional hydrogen atom to discuss the 1D hydrogen atom as a special case taking however the potential $V(x) = -e^2/x$ for $x > 0$ and $V(x) = +\infty$ for $x \leq 0$ invoking this preference for "physical" reasons. Indeed a non degenerate eigenspectrum is "physically" more attractive. In his treatment the twofold degeneracy remains for the potential $V(x) = -e^2/|x|$ for which all the odd solutions of the one dimensional symmetric problem are the only allowed ones correspond to all radial solutions of the 3D problem for $l = 0$. Nieto argues then that by putting an infinite high barrier at the origin all the solutions correspond naturally then to the solutions of the 3D case for l=0.

Moss discussed the hydrogen atom in D dimensions ($D \geq 1$) using the Schrödinger equation and its relativistic counterparts (Klein-Gordon and Dirac equations) [7]. Moss is essentially concerned with the existence of a non degenerate ground state. First, he approaches it using rounded potentials as $x \to \pm 0$ and second, starting out with a dimension $D = 1 + \delta$, letting then at a later stage $\delta \to 0$. In both cases Moss accepts the non degenerate ground state as an existing limit leading to an infinite binding energy [7]. Using the supersymmetric quantum mechanics approach, Imbo and Sukhatme [8] conclude that the odd states of the 1D Coulomb potential are degenerate with the even states and that this potential might be the only potential which violates the non degeneracy theorem in one dimension. Boya et al. [9] argued that the apparent violation of the non degeneracy theorem for bound states in one dimension is due to the fact that the origin separates $\mathbb{R}$ into two regions defined by $(x \leq 0, x \geq 0)$. They reject nevertheless the ground state as a pathological state.

More recently, Núñez-Yépez et al. [10] solved the problem in momentum space. However, they concluded that it is impossible to find eigenfunctions with definite parity for this system which, as we will show below, is the wrong conclusion. The lack of parity of the associated Laguerre polynomials in Eq.(20) of Ref.[10] might have been overlooked and generated the wrong solution in $\mathbb{R}$. Again their arguments led to the wrong results by claiming that the electron is either found on the right hand side or left hand side of the origin, treating the potential as an impenetrable barrier. Martinez et al. [11] concluded that the infinite ground state energy does not exist, however that a twofold degeneracy in the eigenspectrum does. To get more insight into this problem, Davtyan et al. [12] used a Fourier approach technique obtaining however eigenfunctions of even and odd parity, i.e. again a degenerate set of solutions. Their merit nevertheless, was to show that the ground state wave function becomes zero identically and therefore lacks of any physical meaning.

The long and disturbing history of the one dimensional hydrogen atom was examined once again by Gordeyev and Chhajlany [13] seeking solutions by means of a generalized Laplace transform approach to study and unravel the elusive aspects of this eigenvalue problem. Well aware of the problems encountered with the Schrödinger 1D hydrogen atom, their in-depth analysis includes a semi-classical treatment of the problem, obtaining only even energy levels. Their, to some extent, cumbersome mathematical treatment finally yielded a solution differing from all the above authors in one or another way and has clarified some physical aspects of this particular problem. While analyzing the matching of solutions at the origin, Gordeyev et al. recognize the need of proper boundary conditions at $x=0$ to achieve a self-adjoint extension to the Hamiltonian which the authors do not carry out any further. It should be emphasized that no consensus concerning this problem could be claimed, including the results of Gordeyev et al. [12] who eventually revealed contradictory conclusions and errors from previous authors. Moreover, discussion of important previous work on this topic was omitted by these authors as for example the paper of Nuñez et al. which was not mentioned.

Another important contribution is the one by Fisher et al [14] who gave an extensive list of references. Using a functional-analytic approach (compared to a functional-integral approach), Fisher et al. allow different boundary conditions to be imposed on the wave functions $\varphi(x) \in L^2(\mathbb{R})$ at the origin $x=0$ of the real line $\mathbb{R}$. In the case of an attractive Coulomb potential, they report a twofold degenerate Balmer spectrum. One notices that up to a normalization factor their wave functions are those obtained by Nuñez et al.[10]. Nevertheless, Nuñez et al's interpretation differ from the one given by Fisher et al. in the sense that the authors of Ref.[10] obtain a non degenerate energy spectrum while locking the electron in either the positive or negative half line of $\mathbb{R}$, invoking that the 1D Coulomb potential acts as an impenetrable barrier at the origin.

More recently Kurasov analyzed the mathematical properties of the one dimensional Hamiltonian $-d^2/dx^2 - \gamma/x$ [15]. This operator is then compared with the Schrödinger operator containing the 1D Coulomb potential $-\gamma/|x|$. He demonstrates that the Hamiltonian is symmetric but not self-adjoint in the Hilbert space $L^2(\mathbb{R})$ and recognizes that the distribution $(-d^2/dx^2 - \gamma/|x|)\psi$ is in $L^2(\mathbb{R})$ only if the wave function $\psi$ and its first derivative are continuous at the origin. Nevertheless the author speaks about the 'non-penetrability' of this later potential and does not solve the physical problem. Kurasov's analysis was picked up shortly after by Fischer et al. [16] who pointed out that the self-adjoint operators constructed by Kurasov, though correctly constructed, are not necessarily the only self-adjoint operators which can be associated to the 1D Coulomb problem. In Ref. [14] Fischer et al. find a four-parameter family of self-adjoint extensions of the Hamiltonian corresponding to the 1D hydrogen problem. They recognize that different self-adjoint extensions correspond to different boundary conditions imposed upon the wave function $\varphi \in L^2(\mathbb{R})$ at the origin of the real axis. Using the functional-integral approach, only one member of the four-parameter family remains, i.e. the one corresponding to the Dirichlet condition $\varphi(0) = 0$. This finally yields the wrong twofold degenerated energy eigenspectrum and the real line split into a negative and positive half by the singular origin. Their merit was undoubtedly to recognize that mathematics alone will not give the answer to the physical problem since there is no way tell a priori which element of the four-parameter family of Hamiltonians should be chosen in a particular situation. The functional-integral approach unfortunately sifted only one Hamiltonian out of a set of four.

To the best of our knowledge, the latest reference is Grosche and Steiner [17] who is reporting the eigenvalues and eigenfunctions for the "Pure Coulomb Potential in One Dimension" in their Handbook of Feynman Path Integrals for bound states and scattering states. Grosche and Steiner cite Fisher et al. [14] but refer only to the solutions for the positive half line in $\mathbb{R}$. They do not cite Gordeyev et al's contribution [13]. Disturbingly enough, Gordeyev et al's paper does not cite the work of Fisher et al [14]. which leaves one with the impression that the existent controversy of this peculiar challenge had been rather reignited then settled at once.

In this paper, we have reanalyzed the problem and employ a direct unambiguous approach which uses a Fourier transform method combined with some fundamental concepts found in self-adjoint extensions of symmetric operators to obtain the solutions of the one dimensional hydrogen atom. Our approach is a straightforward constructive bottom-up technique. The Fourier transform is the natural integral transformation since the coordinate and momentum spaces play an equivalent role in Schrödinger's picture of quantum mechanics. The Hilbert space, spanned by square integrable eigenfunctions, guarantees the existence of the Fourier transform. Because of the probabilistic interpretation of quantum mechanics, wave functions are square integrable, i.e. $\Psi \in L^2(\mathbb{R})$ and hence the natural correspondence between coordinate space and momentum space is precisely the Fourier transform.

Physically the Schrödinger equation has well defined solutions for a large class of meaningful potentials unless singular points appear. This rather subtle yet essential point is related to the fact that the Hamiltonian $H$ of the 1D hydrogen atom is hermitian but not self-adjoint [18]. Therefore one has to look for its self-adjoint extensions which amounts to specifying some particular boundary conditions to describe the behaviour of the wave function at the origin. Different self-adjoint extensions correspond to different quantum physical situations [19] and hence appropriate matching conditions of the wave function must be chosen at the origin. It is important to recognize that self-adjointness is a mathematical property of hermitian operators which does not necessarily describe a particular physical situation, but it is rather the physics of the underlying problem which dictates the self-adjoint extension required by the problem. This is the key point for the 1D hydrogen atom which leads Nuñez et al [10] to wrong results (conclusions), i.e. confining the electron either to the region $x \leq 0$ or the region $x \geq 0$ and hence arriving to the wrong interpretation. For instance, the matching of solutions at singular points plays a fundamental role such as found in problems with delta distribution potentials where the first derivative of the wave function is discontinuous at the singularity, leading to the discussion of resonances and the law of exponential decay in quantum mechanics [20].

We explicitly found eigenfunctions both in the momentum and coordinate spaces for the 1D hydrogen atom. No twofold degeneracy was observed while the complete Balmer energy eigenspectrum identical to the 3D counterpart was recovered in spite of the fact that only odd eigenfunctions are allowed. Moreover, we showed that the nodeless ground state with infinite negative energy does not exist.

## 2.    The one dimensional hydrogen atom.

The one-dimensional hydrogen atom pursues the solution to the problem of an electron moving in the one dimensional potential

$$V(x) = -\frac{e}{|x|} \tag{1}$$

The Schrödinger equation for this system is therefore

$$-\frac{\hbar^2}{2m}\frac{d^2\Psi}{dx^2} - \frac{e^2}{|x|}\Psi = E\Psi \tag{2}$$

Let us introduce the real dimensionless quantity $\alpha$ by putting

$$E = \frac{-\hbar^2}{2ma_0^2 \alpha^2} \tag{3}$$

where $a_0$ is the Bohr radius. After changing to a new independent variable $z$ defined as

$$z = \frac{2x}{\alpha a_0} \tag{4}$$

we obtain the differential equation

$$\frac{d^2\Psi}{dz^2} - \frac{1}{4}\Psi + \frac{\alpha}{|z|}\Psi = 0 \tag{5}$$

At this point, we shall use a novel method based on the Fourier transform method to find the eigenvalues and eigenfunctions.

By considering the asymptotic behaviour of the solution of Eq. (5) for $|z| \ll 1$ and $|z| \gg 1$ respectively, we use the following Ansatz

$$\Psi(z) = z\, e^{\pm z/2}\, \Phi(z) \tag{6}$$

where $\Phi(z)$ is a regular function at the origin and $\lim_{z \to 0} \Phi(z) = const$. Since we are only interested in bound states, we require that $\Psi(z)$ be a square integrable function, i.e. $\Psi(z) \in L^2(\mathbb{R})$. We shall now address some aspects of our strategy concerning the sign used in the Ansatz (6) which is essential to the method. Contrary to the standard procedure, we have to choose the "$+$" sign for $z > 0$ and vice versa the "$-$" sign for $z < 0$ such that our function $\Phi(z)$ is no longer a polynomial but a square integrable function which is regular at the origin and hence a square integrable function, i.e. $\Phi \in L^2(\mathbb{R})$. Hence, we obtain the following differential equation for the function $\Phi(z)$

$$z\Phi'' + (2 \pm z)\Phi' + (\alpha\, \mathrm{sgn}(z) \pm 1)\Phi = 0 \tag{7}$$

Now we use the Plancherel theorem [19] stating that the Fourier transform extends uniquely to a unitary map of $L^2(\mathbb{R})$ onto $L^2(\mathbb{R})$ to write Eq. (7) in terms of the Fourier transform of $\Phi(z) \in L^2(\mathbb{R})$

$$(-ik^2 \mp k)\widetilde{\Phi}'(k) = -\alpha\, \mathrm{sgn}(z)\, \widetilde{\Phi}(k) \tag{8}$$

which can then be solved directly by variable separation yielding

$$\widetilde{\Phi}(k) = A\left(\frac{k}{k \mp i}\right)^{\pm \alpha\, \mathrm{sgn}\, z} \tag{9}$$

The existence of the inverse transform of $\widetilde{\Phi}(k)$ is also guaranteed by the Plancherel theorem. Therefore we obtain from Eqs. (6) and (9)

i)      for $z > 0$      $$\Psi_I(z) = A\, z\, e^{z/2} \int \frac{dk}{2\pi}\, e^{ikz} \left(\frac{k}{k - i}\right)^{\alpha} \tag{10a}$$

ii)  for $z < 0$

$$\Psi_{II}(z) = B\, z\, e^{-z/2} \int \frac{dk}{2\pi} e^{ikz} \left( \frac{k}{k+i} \right)^{\alpha}  \qquad (10b)$$

Now we use the general requirement that the wave function $\Psi(z)$ must be single-valued [21], which holds if and only if the exponent $\alpha$ in the above expression (10a and 10b) is an integer, i.e. $n \in \mathbb{N}$.

The condition that the wave function has to be single-valued leads naturally to the condition $\alpha = n \in \mathbb{N}$ which hence implies that the integrands of Eqs. (10a) and (10b) have poles of order $n$ at $k = i$ and $k = -i$ respectively. The integral defining our function $\Psi(z)$ can consequently be computed using the residuum theorem. Defining

$$F_{\pm}(z) = \int \frac{dk}{2\pi} e^{ikz} \left( \frac{k}{k \mp i} \right)^n \qquad (11)$$

a straightforward calculation yields

$$F_{+}(z) = \frac{1}{(n-1)!} \frac{d^n}{dz^n}(z^{n-1} e^{-z}) \qquad (12)$$

We are now able to explicitly write down the wave function after recalling the following definition of the associated Laguerre polynomials $L_n^m(z)$ [22]

$$L_n^m(z) = \frac{d^m}{dz^m}\left\{ e^z \frac{d^n}{dz^n}(z^n e^{-z}) \right\}$$

and evaluating them for $m = 1$, we obtain respectively

for $z > 0$ $\qquad \Psi_n(z) = A\, z\, e^{-z/2} L_n^1(z) \qquad (13a)$

for $z < 0$ $\qquad \Psi_n(z) = B\, z\, e^{z/2} L_n^1(-z). \qquad (13b)$

Using Eq. (3) for $\alpha = n \in \mathbb{N}$, the quantized energy levels of the one dimensional hydrogen atom are now given by

$$E_n = -\frac{\hbar^2}{2ma_0^2} \frac{1}{n^2} \qquad (14)$$

## Solution by means of confluent hypergeometric functions.

We recognize Eq.(5) as Whittaker's form of the confluent hypergeometric equation [23]. Power series solutions can always be found using the Frobenius method [24]. For $z > 0$ our differential equation (7)

$$z\Phi'' + (2-z)\Phi' + (\alpha-1)\Phi = 0$$

is solved by the confluent hypergeometric function (Kummer's function) regular at the origin

$$\Phi(z) = {}_1F_1(1-\alpha, 2, z)$$

which is the regular solution of the differential equation for the associated Laguerre polynomials $L_n^m(z)$ for $m=1$ and $\alpha = n \in \mathbb{N}$

$$zw'' + (1+m-z)w' + (n-m)w = 0$$

and therefore we can conclude that

$$L_n^1(z) = C_n \, {}_1F_1(1-n, 2, z)$$

with $C = n!$, where the associated Laguerre polynomial $L_n^1(z)$ can be written as

$$L_n^1(z) = n e^z \frac{d^n}{dz^n}(z^{n-1} e^{-z})$$

The solution for the case $z < 0$ is obtained from the differential equation

$$z\Phi'' + (2+z)\Phi' + (-\alpha+1)\Phi = 0$$

which is obtained from the former equation by means of the reflection $z \to -z$. In this case the solution is

$$\Phi(z) = L_n^1(-z) = C \, {}_1F_1(1-n, 2, -z)$$

The second solution of Kummer's differential equation is not regular at the origin and therefore should be discarded.

Since the solutions (13a) and (13b) have to be matched at the origin, Schrödinger's equation (5) is integrated in the interval $[-\varepsilon, +\varepsilon]$ and the limit $\varepsilon \to 0$ has to be taken

$$\Psi'(\varepsilon) - \Psi'(-\varepsilon) + n\left(-\int_{-\varepsilon}^{0} dz \, \frac{\Psi(z)}{z} + \int_{0}^{\varepsilon} dz \, \frac{\Psi(z)}{z}\right) = 0 \qquad (15)$$

From Eqs (13a) and (13b), we notice that $\frac{\Psi(z)}{z}$ is regular and continuous at the origin $z=0$ and hence both integrals of Eq. (15) vanish after taking the limit $\varepsilon \to 0$ and therefore we conclude that the first derivative must be continuous at $z=0$. It follows that $A=B$ in (13a) and (13b). Hence, we can write the solutions as

$$\Psi_n(z) = A_n \, z \, e^{-|z|/2} \, L_n^1(|z|) \tag{16}$$

where $\Psi_n(z)$ is a real function up to a constant phase.

The normalization constant $A_n$ of our solutions (16) becomes then (see Appendix A)

$$A_n = (-)^n \frac{\sqrt{2na_0}}{n! n^3 a_0^2} \frac{na_0}{2}. \tag{17}$$

In momentum space the normalized wave function becomes

$$\varphi_n(k) = \sqrt{\frac{a_0 n}{\pi}} \frac{1}{1+n^2 a_0^2 k^2} \left\{ \left( \frac{1-ina_0 k}{1+ina_0 k} \right)^n - \left( \frac{1+ina_0 k}{1-ina_0 k} \right)^n \right\} \tag{18}$$

which is also an odd parity function of $k$. From this expression one sees that when $n=0$, there is no solution since the wave function vanishes everywhere and hence does no longer satisfy Schrödinger's equation (see discussion below). Finally, it can be easily shown that the Parseval-Plancherel relation

$$\int_{\mathbb{R}} dk \, |\varphi(a_0 k)|^2 = \int_{\mathbb{R}} dx \, |\Psi(x)|^2 \tag{19}$$

is satisfied as expected (see Appendix A).

From Eq.(16) it follows that the wave functions have odd parity under spatial reflections. The boundary condition of continuity of the first derivative at the origin excludes the even eigenfunctions. Parity requirements stem from the fact that the parity operator $P$ commutes with the Hamiltonian $H$, i.e. $[H,P]=0$. Nevertheless due to the singularity at the origin, the matching conditions of the wave function from both sides of the singularity must be computed by integrating the Schrödinger equation which automatically rules out the even parity solutions.

Thus the complete Balmer energy spectrum (14) is obtained and consequently we conclude that the levels of the one dimensional hydrogen atom are not degenerate in agreement with the theorem that stipulates that one-dimensional systems cannot have degenerate states. Nevertheless, as correctly pointed out by Loudon [2], the non degeneracy theorem does not necessarily have to be valid for a potential with singular points. Despite the fact that the former statement is correct, one has to be careful when matching the solutions around the singular points of the potential before claiming a possible violation of the non degeneracy theorem.

Instead of working in coordinate space, Nuñez et al. [10] have derived the following solution for the one dimensional hydrogen problem in momentum space

$$\varphi_{\pm}(k) = \frac{A_{\pm}\alpha^2}{1+\alpha^2 k^2}\left(\frac{1 \mp i\alpha k}{1 \pm i\alpha k}\right)^{\alpha} \qquad (20)$$

where $\varphi_{\pm}(k)$ refers to the solution in momentum space corresponding to the solution in coordinate space for $z > 0$ and $z \leq 0$ respectively. Again, one sees immediately that $\alpha = n \in \mathbb{Z}$, i.e. that $\alpha$ must be an integer such that the Fourier transform of the wave function $\varphi_{\pm}(k)$ is single-valued. This leads then to the Balmer series for the eigenvalues $E_n = -e^2/(2a_0 n^2)$. In Ref. [10] the authors argued that since "the singularity of the Coulomb potential at the origin acts as an impenetrable potential barrier", the electron remains confined in either one of the two regions defined by $z > 0$ and $z \leq 0$. This claim leads then to their solution (20a) $\psi_+^n(x)$ and (20b) $\psi_-^n(x)$ respectively. Moreover, these solutions do not fulfil Schrödinger's equation. This is easily understood since the first derivative is not continuous at the origin, which would as well lead to a delta potential at the origin.

Now we like to find out if the singularity at the origin $x = 0$ results in an impenetrable barrier confining the electron either to the left or right hand side of the origin on the real axis. As can be seen in Appendix B the probability density current vanishes at the origin since the eigenstates given by Eq. (16) vanish at that point. This observation leads the authors of Ref [10] to conclude that the electron is either found in $x < 0$ or $x > 0$. Nevertheless we disagree with this conclusion because the potential is attractive (well potential and not a barrier). This can be illustrated for example with the classic example of the attractive potential $V(z) = -c\,\delta(z)$ [25] yielding one single bound state or simply with a finite square well which yields at least one bound state. The probability of finding a particle in the classically forbidden regions of the aforementioned potentials is non zero. Nevertheless the probability density current $j(z)$ is zero at the edges of the wells when considering only bound states. In spite of this fact, particles described by scattering states can cross the potential such that it cannot be concluded that the potential well acts as an impenetrable barrier if the probability density current for bound states vanishes at a particular point. This rather subtle point seems to have contributed to some confusion regarding the interpretation of the particular problem.

We can use a semi-classical picture to argue that an electron moving under the influence of the 1D Coulomb potential is not confined to remain either to the right or left of the origin. In fact the velocity of the electron with negative energy (bound state) will be given by $v(x) = \sqrt{2/m(\alpha/|x| - |E|)}$, which implies that the velocity increases while coming closer to $x = 0$ and finally diverges at this point. Let $\Delta t_0$ be the time the electron spends in a space interval of width $\Delta$ which includes the origin and $\Delta t$ the time the electron spends in an interval of same width excluding the origin. Then it is easy to see that $\lim_{\Delta \to 0} \frac{\Delta t_0}{\Delta t} = 0$ and hence it follows that the probability to find the electron in an interval around the origin is zero compared to any other interval of the same range.

At this point, a few words concerning the paper of Gordeyev et al. [13] have to be added to our discussion. The contribution of Gordeyev and Chhajlany consists in a thorough analysis of the 1D hydrogen atom, eventually due to the controversial results in the literature. Using semi-classical arguments these authors arrive at force acting upon the electron which provides harmonic oscillations about the point of stable equilibrium. The presence of a repulsive term expressed as a delta distribution function makes "*the force vanish at the origin, but does not forbid a penetration of the electron through the origin*". As the authors correctly point out, this is contradicting the observations from Andrews [3] and, from our point of view, the comments of Nuñez et al [10], stipulating that an attractive potential can act as an impenetrable barrier in quantum mechanics.

### 3. The self-adjoint extension of the Hamiltonian.

Now we shall appeal to the important aspect related to the fundamental postulate of quantum mechanics stipulating that each observable is represented as a self-adjoint operator acting on the Hilbert space associated to the considered physical system. As mentioned in the introduction the Hamiltonian $H$ of the 1D hydrogen atom defined by Eq. (2) is hermitian (or symmetric) but not self-adjoint. Its self-adjoint extensions have been thoroughly discussed in ref. [14] and a four-parameter family of extensions was found. Each of these extensions corresponds to a set of different boundary conditions to be imposed on the wave function $\varphi \in L^2(\mathbb{R})$ at the origin of the real axis.

In Appendix B we shall determine a self-adjoint extension of our Hamiltonian which is compatible with the boundary conditions of continuity of the first derivative at the origin expressed by Eq. (15). Let us emphasize again that the physics of a problem will define the adequate self-adjoint extension compatible with the required boundary conditions.

Our conclusion in the present case is that the hermitian Hamiltonian defined on the one-parameter domain $D(H_R)$ with

$$D(H_R) = \left\{ \varphi \in L^2(\mathbb{R}) : \varphi, \varphi' \in \mathcal{AC}_{loc}(\mathbb{R} \setminus \{0\}), (\frac{d^2}{dz^2} - \frac{1}{4} + \frac{\alpha}{|z|})\varphi \in L^2(\mathbb{R}), \right.$$

$$\left. \begin{pmatrix} \lim_{z \uparrow 0} \varphi(z) \\ \lim_{z \uparrow 0} \varphi'(z) \end{pmatrix} = R \begin{pmatrix} \lim_{z \downarrow 0} \varphi(z) \\ \lim_{z \downarrow 0} \varphi'(z) \end{pmatrix} \right\} \quad (21)$$

where $R$ is the $2x2$ rotation matrix

$$R = \begin{pmatrix} \cos\theta & -\sin\theta \\ \sin\theta & \cos\theta \end{pmatrix}$$

and $\mathcal{AC}_{loc}(\Lambda)$ denotes the set of complex-valued functions $\varphi$ on $\Lambda \subseteq \mathbb{R}$ which are absolutely continuous on every compact subset of $\Lambda$ and $\varphi'$ and $\varphi''$ have to be defined Lebesgue almost everywhere, is indeed self-adjoint. Moreover, the eigenfunctions whose first derivatives are continuous at the origin (Eq. (15)) are contained in $D(H_R)$ for the special case $\theta = 0$ (see Appendix B).

## 4. Infinite negative energy ground state ?.

Loudon [2] proposed a formal solution for the ground state of the one dimensional hydrogen atom by taking the limit $\alpha \to 0$ in Eqs.(3) and (5). In this case, the ground state $E_0 = -\infty$ and the corresponding nodeless eigenfunction which is a solution of Eq.(5) becomes

$$\Psi(z) = \frac{1}{\sqrt{2}} e^{-|z|/2}$$

Notice however, that the variable substitution given by (4) is singular in the limit $\alpha \to 0$. Moreover, starting out with Eq.(2) and expressing the wave function $\Psi$ as a function of $x$ using (4), one immediately sees that the ground state wave function does not satisfy Schrödinger's equation (2) and hence has to be discarded at once as a singular improper limit.

## 5. Conclusions

The peculiar features of this one-dimensional quantum system have led several authors to similar results, discrepancies, some contradictions and confusion. It is rather curious that until now no consensus had been reached over decades concerning the intricacies of the mathematical and physical aspects of this particular problem. No concurrence of results could be claimed so far, including Gordeyev and Chhajlany's paper which omitted the discussion of some important work from previous authors. Gordeyev and Chhajlany recognize that the Hamiltonian of this problem is symmetric but not self-adjoint. However they do not develop further this important aspect of the analysis. Though we partially agree with Gordeyev and Chhajlany's results and due to the fact that to the best of our knowledge, the use of generalized Laplace transforms has never been proposed nor applied as a general method to solve eigenvalue problems in quantum mechanics, we believe that, despite their solution of the quantum mechanical 1D hydrogen atom, our method is much more straightforward and transparent due to its general constructive approach which automatically selects the square integrable (physical) solutions of the Schrödinger equation making use of the natural equivalent space or momentum representation of Schrödinger's picture by means of Fourier integral transforms. Moreover, we explicitly compute the adequate self-adjoint extension of the Hamiltonian which is compatible with the boundary conditions and fulfil the requirements of quantum mechanics.

Some final remarks: our 1D problem is fully equivalent to the 3D hydrogen problem for $l=0$ which nullifies the centrifugal potential term $l(l+1)/r^2$. As is well known, the solutions of Schrödinger's hydrogen atom can be formally written as $\Psi_{nlm}(r,\Omega) = c_{nl} R_{nl}(r) Y_{lm}(\Omega)$, where $Y_{lm}(\Omega)$ are the spherical harmonics, $R_{nl}(r)$ are the radial eigenfunctions and $c_{nl}$ is a normalization constant respectively. If we write $R_{nl}(r) = u_{nl}(r)/r$ then $u_{nl}(r)$ satisfies for $l=0$ the same differential equation as the eigenfunction of the 1D hydrogen, for the positive half region $x>0$ which reflects a different topology in the 1D problem. The whole Balmer spectrum is recuperated as well.

We have reanalyzed the subject of the one dimensional hydrogen atom using a general Fourier transform method to find the eigenvalues and eigenfunctions in both momentum and spatial coordinates. Our analysis has showed that the bound states in a 1D Coulomb potential are given by the complete Balmer eigenvalue spectrum and odd parity eigenfunctions due to the matching conditions at the origin which are consistent with the self-adjoint extension of the Hamiltonian $H$ found in Section 3. The nodeless ground state eigenfunction does not satisfy the original Schrödinger equation and consequently no such state with infinite binding energy exists. Finally and in spite of the singularity of the potential at the origin, the validity of the non-degeneracy theorem was corroborated with our results for this special 1D quantum problem. Although the non degeneracy theorem does not necessarily apply when the potential has singularities, one should be particularly cautious when matching solutions at these singular points. Therefore, the one dimensional hydrogen atom is not a pathological example that contradicts the aforementioned theorem.

**Acknowledgements.**

The authors are grateful to partial funding by FONDECYT # 1050266 (G.P.) and DICYT # 040131RB (U.R.). Enlightening discussions with Dr. Martin Porrmann from the II Institute for Theoretical Physics of the University of Hamburg, (Germany) are also gratefully acknowledged. One of us (G.P.) is thankful for the hospitality received at the same Institute in Hamburg (Germany). We like in particular to thank the reviewers of this paper due to their valuable contributions which lead us to discuss the self-adjointness of the Hamiltonian.

## Appendix A

The normalized eigenfunctions of our problem can be written as

$$\Psi_n(x) = \Psi_n^+(x)\Theta(x) + \Psi_n^-(x)\Theta(-x) \tag{A.1}$$

which are defined in the entire domain $\mathbb{R}$ (see Eq. 16), can easily be recovered applying the Fourier transform to the expression $\varphi_n(k) = \varphi_n^+(k) + \varphi_n^-(k)$.

Let us use Eq.(18) to compute $\Psi_n^+(x)$ for $x > 0$:

$$\Psi_n^+(x) = A_n^+ \int_\mathbb{R} \frac{dk}{\sqrt{2\pi}} e^{ikx} \frac{(1 - ina_0 k)^{n-1}}{(1 + ina_0 k)^{n+1}} \tag{A.2}$$

with normalization constant $A_n^+ = \left(\frac{na_0}{\pi}\right)^{1/2}$. The integral (A.2) can be evaluated using the residuum theorem, observing that the integral has a pole of order $n+1$ at $k = i/(na_0)$:

$$\Psi_n^+(x) = \frac{\sqrt{2\pi}\,(-)^{n-1} A_n^+}{n!\, n^2 a_0^2} e^{x/na_0} \frac{d^{n-1}}{dx^{n-1}}\left\{x^n\, e^{-2x/na_0}\right\} \tag{A.3}$$

Using the dimensionless variable $z = (2x)/(na_0)$ (A.3) becomes

$$\Psi_n^+(z) = (-)^n \frac{\sqrt{2\pi}\, A_n^+}{n!\, n^2 a_0^2} \frac{na_0}{2} e^{-z/2}\, e^z \frac{d^{n-1}}{dz^{n-1}}(z^n\, e^{-z}) \tag{A.4}$$

If the following definition of the associated Laguerre polynomials given by the expression 8.970 in Ref [26]:

$$\mathcal{L}_n^m(z) = \frac{1}{n!} e^z z^{-m} \frac{d^n}{dz^n}(e^{-z} z^{n+m}) \tag{A.5}$$

is used, while putting again $z = (2x)/(na_0)$ in (A.4) we obtain :

$$\Psi_n^+(x) = (-)^{n-1} \frac{A_n \sqrt{2\pi}}{n^3 a_0^2} x e^{-x/na_0} \mathcal{L}_{n-1}^1(2x/na_0) \tag{A.6}$$

From the alternate definition of the associated Laguerre polynomials $L_n^m(z)$ [25] we obtain the identity :

$$L_n^1(z) = -n! \mathcal{L}_{n-1}^1(z) \tag{A.7}$$

The normalized eigenfunctions $\Psi_n^\pm(x)$ (A.1) of our problem are then given by

$$\Psi_n^+(x) = (-)^n \frac{\sqrt{2na_0}}{n! n^3 a_0^2} x e^{-x/(na_0)} L_n^1(2x/na_0) \quad x > 0 \tag{A.8}$$

and $\Psi_n^-(x) = -\Psi_n^+(-x)$ for $x < 0$. The wave function $\Psi_n(x)$ can then be recovered from (A.1).

Using the following identity (see 21-7-17 of Ref.[27]) :

$$\int_0^\infty dz\, e^{-z} z^{m+1} [L_n^m(z)]^2 = \frac{(2n-m+1)}{(n-m)!}(n!)^3 \tag{A.9}$$

One can immediately prove using (A.8) that

$$\int_{-\infty}^{+\infty} dx\, |\Psi_n(x)|^2 = 1$$

which corroborates that Eq.(19) is the corresponding normalized eigenfunction in momentum space.

## Appendix B.

**Proof of self-adjointness of $H_R$.**

Let us consider the differential operator of Eq. (2)

$$H = \frac{d^2}{dz^2} - \frac{1}{4} - \frac{\alpha}{|z|} \tag{B1}$$

which is defined on the domain of arbitrary often differentiable complex-valued functions with compact support in $\mathbb{R}$ excluding the origin $\mathbb{C}_0^\infty(\mathbb{R}\setminus 0)$. The aim of this appendix is to find a dense domain in the Hilbert space $L^2(\mathbb{R})$ on which $H$ is self-adjoint.

The Hamiltonian $H$ defined in (B1) is hermitian (or symmetric). Let $D(H)$ be the domain of the Hamiltonian. Thus for any $\varphi \in D(H)$, an integration by parts using the principal-value prescription shows that:

$$<\varphi, H\varphi> - <H^*\varphi, \varphi> = \lim_{\varepsilon \to 0}\left[\frac{2i}{\lambda}\right]\{j(-\varepsilon) - j(\varepsilon)\} = 0 \tag{B2}$$

where we have used the abbreviation $\lambda = 2me^4/\alpha^2\hbar^3$ and where $j(z)$ is the probability current density defined by

$$j(z) = \frac{\lambda}{2i}\left[\overline{\varphi}(z)\varphi'(z) - \overline{\varphi}'(z)\varphi(z)\right] \tag{B3}$$

We conclude from Eq. (B2) that the probability current density must be continuous at the origin. From a physics point of view, this condition means that the probability current incident from minus infinity towards the origin is equal to the one flowing out from the origin towards plus infinity. The expression given by Eq.(B3) is not linear in the wave function and hence it is not possible to define a linear subspace which fulfils the condition given by Eq. (B2). However, in order to guarantee the superposition principle of quantum mechanics, we shall look for a linear condition to be imposed upon the wave function which implies the continuity condition of the probability current. There are many of such linear conditions [14]. Indeed, the problem of choosing the "right" self-adjoint extension is not a mere mathematical task but is intimately related to the physics of the problem [18]. One linear condition which guaranties the continuity of the probability current density is defined by the one-parameter relation

$$\begin{pmatrix} \lim_{\varepsilon\uparrow 0}\varphi(\varepsilon) \\ \lim_{\varepsilon\uparrow 0}\varphi'(\varepsilon) \end{pmatrix} = R \begin{pmatrix} \lim_{\varepsilon\downarrow 0}\varphi(\varepsilon) \\ \lim_{\varepsilon\downarrow 0}\varphi'(\varepsilon) \end{pmatrix} \tag{B4}$$

where $R$ is the unitary (rotation) 2x2 matrix

$$R = \begin{pmatrix} \cos\theta & -\sin\theta \\ \sin\theta & \cos\theta \end{pmatrix} \quad (B5)$$

Now we apply this linear condition to the wave function and its derivative to define the extended operator $H_R$ whose domain becomes:

$$D(H_R) = \left\{ \varphi \in L^2(\mathbb{R}) : \varphi, \varphi' \in \mathcal{AC}_{loc}(\mathbb{R}\setminus\{0\}), \left(\frac{d^2}{dz^2} - \frac{1}{4} + \frac{\alpha}{|z|}\right)\varphi \in L^2(\mathbb{R}), \right.$$

$$\left. \begin{pmatrix} \lim_{z\uparrow 0}\varphi(z) \\ \lim_{z\uparrow 0}\varphi'(z) \end{pmatrix} = R \begin{pmatrix} \lim_{z\downarrow 0}\varphi(z) \\ \lim_{z\downarrow 0}\varphi'(z) \end{pmatrix} \right\} \quad (B6)$$

Here $\mathcal{AC}_{loc}(\Lambda)$ denotes the set of complex-valued functions $\varphi$ on $\Lambda \subseteq \mathbb{R}$ which are absolutely continuous on every compact subset of $\Lambda$, and $\varphi'$ and $\varphi''$ are understood to be defined Lebesgue almost everywhere. Next we shall proof that the operator

$$H_R : D(H_R) \to L^2(\mathbb{R}) \quad (B7)$$

with $(H_R \varphi)(z) =: \varphi''(z) - \frac{1}{4}\varphi(z) - \frac{\alpha}{|z|}\varphi(z)$ is self-adjoint for each value of $\theta$.

In fact, let us choose $\varphi \in D(H_R)$ and $\phi \in D(H_R^*)$. Integration by parts leads to the relation

$$<\phi, H_R \varphi> - <H_R^* \phi, \varphi> = \lim_{\varepsilon \uparrow 0}\left[\overline{\phi}(\varepsilon)\varphi'(\varepsilon) - \overline{\phi}'(\varepsilon)\varphi(\varepsilon)\right]$$
$$- \lim_{\varepsilon \downarrow 0}\left[\overline{\phi}(\varepsilon)\varphi'(\varepsilon) - \overline{\phi}'(\varepsilon)\varphi(\varepsilon)\right] \quad (B8)$$

Using now the fact that $\varphi \in D(H_R)$, we obtain for the left hand side of (B8)

$$l.h.s. = \lim_{\varepsilon\downarrow 0}\varphi(\varepsilon)\left[\sin\theta(\lim_{\varepsilon\uparrow 0}\overline{\phi}(\varepsilon) - \cos\theta(\lim_{\varepsilon\uparrow 0}\overline{\phi}'(\varepsilon) + (\lim_{\varepsilon\downarrow 0}\overline{\phi}'(\varepsilon))]\right]$$
$$- \lim_{\varepsilon\downarrow 0}\varphi'(\varepsilon)\left[-\sin\theta(\lim_{\varepsilon\uparrow 0}\overline{\phi}'(\varepsilon) - \cos\theta(\lim_{\varepsilon\uparrow 0}\overline{\phi}(\varepsilon) + (\lim_{\varepsilon\downarrow 0}\overline{\phi}(\varepsilon))]\right] \quad (B9)$$

Since $H_R$ is a symmetric extension of $H$, expression (B9) must vanish identically for all $\phi \in D(H_R^*)$. From relation (B9) it follows that

$$\begin{pmatrix} \lim_{\varepsilon\uparrow 0}\phi(\varepsilon) \\ \lim_{\varepsilon\uparrow 0}\phi'(\varepsilon) \end{pmatrix} = \begin{pmatrix} \cos\theta & -\sin\theta \\ \sin\theta & \cos\theta \end{pmatrix} \begin{pmatrix} \lim_{\varepsilon\downarrow 0}\phi(\varepsilon) \\ \lim_{\varepsilon\downarrow 0}\phi'(\varepsilon) \end{pmatrix} \quad (B10)$$

But this condition implies that $\phi \in D(H_R)$. Since $D(H_R) \subset D(H_R^*)$, we therefore conclude that $D(H_R) = D(H_R^*)$, i.e. $H_A$ is a self-adjoint extension of $H$ for each value of $\theta$. In particular, if we choose $\theta = 0$, we obtain the physical condition of continuity of the first derivative of the wave function at the origin found in Eq.(15) of section 2.